\documentclass[aps,
prd,
reprint,
amsmath,
amssymb,
amsthm, 
bm,
]{revtex4-2}

\usepackage{graphicx}
\usepackage{dcolumn}
\usepackage{multirow} 
\usepackage{empheq}
\usepackage{xcolor}
\usepackage{soul}
\usepackage{hyperref}
\usepackage{placeins} 
\usepackage{bm} 
\usepackage{amsmath}
\usepackage[normalem]{ulem}

\renewcommand\hl[1]{#1}

\begin{document}


\title{Fully data-driven time-delay interferometry with time-varying delays}



\author{Quentin Baghi}
\email[]{quentin.baghi@cea.fr}

\affiliation{IRFU, CEA, Universit\'{e} Paris-Saclay, F-91191 Gif-sur-Yvette, France}

\author{John G. Baker}
\author{Jacob Slutsky}
\author{James Ira Thorpe}

\affiliation{Goddard Space Flight Center \\
	Mail Code 663\\
	8800 Greenbelt Rd, Greenbelt, 
	Maryland 20771, USA}


\date{September 16, 2022}

\begin{abstract}
	\hl{Raw space-based gravitational-wave data like LISA's phase measurements are dominated by laser frequency noise. The standard technique to make this data usable for science is time-delay interferometry (TDI), which cancels laser noise terms by forming suitable combinations of delayed measurements. We recently introduced the basic concepts of an alternative approach which, unlike TDI, does not rely on independent knowledge of temporal correlations in the dominant noise.} Instead, our automated Principal Component Interferometry (aPCI) processing only assumes that one can produce some linear combinations of the temporally nearby regularly spaced phase measurements, which cancel the laser noise. Then we let the data reveal those combinations. 
	Our previous work relies on the simplifying additional assumption that the filters which lead to the laser-noise-free data streams are time-independent. In LISA, however, these filters will vary as the constellation armlengths evolve. Here, we discuss a generalization of the basic aPCI concept compatible with data dominated by a still unmodeled but slowly varying noise covariance. Despite its independence on any model, aPCI successfully mitigates laser frequency noise below the other noises' level, and its sensitivity to gravitational waves is the same as the state-of-the-art second-generation TDI, up to a 2\% error.
\end{abstract}


\maketitle


\section{\label{sec:intro}Introduction}

Observing gravitational waves (GWs) with heterodyne interferometric detection cannot be done without cancelling the overwhelming noise stemming from the stochastic frequency fluctuations of the current technology lasers. For space-based interferometers like LISA, this crucial operation is performed on ground, once the data is downloaded from the satellites to Earth. The standard and most fully  developed method to achieve this cancellation is time-delay interferometry (TDI)~\cite{tinto_cancellation_1999, tinto_time-delay_2020} \hl{a post-processing technique which performs adequate combinations of delayed phase measurements that represent virtual multiple beam interferometers}~\cite{vallisneri_2005, muratore_revisitation_2020}, to nearly nullify laser noise. To prepare the data analysis of the the Laser Interferometer Space Antenna (LISA)~\cite{danzmann_lisa_2017}, significant efforts have been made to assess the performance and underlying characteristics of TDI (see \cite{tinto_time-delay_2020} and references therein), including studying its interplay with anti-aliasing filters~\cite{bayle_effect_2019}, measurements units~\cite{bayle_adapting_2021}, clock jitters~\cite{hartwig_clock-jitter_2020} and clock synchronization~\cite{hartwig_time_2022}, new noise-cancelling combinations~\cite{muratore_revisitation_2020} and the construction of null channels~\cite{muratore2022, muratore2022b}.

A new approach to the laser frequency noise problem has gained interest in the last two years, which formulates how the noise enters phase measurements with a design matrix, and interprets TDI as the solution to a linear algebra problem. Romano and Woan~\cite{romano_principal_2006} took a first step in that direction (further explored in \cite{leighton_principal_2016}) by showing that we can derive TDI variables from the eigenvectors of the laser noise covariance matrix, using a simple toy model in the time domain. More recently, we formalized this idea in the frequency domain, an approach that we named principal component interferometry (PCI), for which we provided first evidences for its suitability to parameter inference~\cite{baghi_statistical_2021}. Alternatively, authors in Ref.~\cite{vallisneri_tdi-infinity_2020} defined the TDI combinations from the null space of the design matrix itself. Another group~\cite{tinto_matrix_2021} demonstrated the equivalence between this technique and the algebraic definition of TDI as a ring over the space of polynomials in delay operators, bridging the matrix-based approaches with \hl{earlier studies of TDI based on set theory}~\cite{vinet_algebraic_2002, nayak_algebraic_2004}. 

In a recent work~\cite{baghi_model-independent_2021} (that we refer to as Paper I in the following), we proposed to further benefit from the power of matrix representation by directly analysing the interferometric measurements without assuming any prior knowledge on their correlations, except that they must extend further than a minimal time set by the problem timescale. The method, called aPCI for ``automated PCI", first forms a data matrix from replicas of phase measurements shifted by a integer number of samples backward or forward in time. Then, performing the matrix's principal component analysis (PCA) yields an array of components ordered by their variance, where the lowest variance components are almost free from laser noise. This handful of variables are sensitive to GWs, and can be used for source detection and characterisation. The process can be understood as a multivariate version of singular spectrum analysis (SSA), a technique broadly used in signal processing (see, e.g., \cite{Golyandina2013}). We proved this concept in the case of constant (but unequal) interferometric links, demonstrating that aPCI's sensitivity is virtually the same as first-generation TDI, i.e., TDI combinations tailored for fixed armlengths. 

In this work, we present an upgrade of the aPCI method suitable for time-varying links, making it applicable to realistic space-borne measurements like the phase-meter and auxiliary system telemetry that will be delivered by the satellites in LISA's time-evolving constellation. This new version is very similar to the time-independent aPCI, except that additional columns are appended to the data matrix to account for polynomial time variations. \hl{It directly compares to second-generation TDI, which is the extension of first-generation TDI that removes laser noise up to linear effects in the armlenghts of a flexing constellation.}
In Section~\ref{sec:theory}, we recall aPCI's theoretical foundations and present its time-varying extension to arbitrary order in time. Then, in Section~\ref{sec:laser_noise_mitigation} we apply the first-order version of the method to numerical simulations of LISA's interferometer data featuring a flexing constellation and show that the method successfully \hl{mitigates} laser frequency noise. In Section~\ref{sec:sensitivity}, we compute the first-order aPCI response to both instrumental noise and GW signals, which allows us to derive its sensitivity and compare it to standard TDI. Finally, in Section~\ref{sec:discussion} we discuss the implications of our findings and outline further developments needed to strengthen aPCI's robustness for real-world data analysis.

\section{\label{sec:theory}Theoretical framework}

Space-based gravitational-wave detectors will send to the ground several interferometers output data, which can be expressed in relative frequency deviations. Similarly as in Ref.~\cite{bayle_adapting_2021}, we denote the corresponding measurements as $y_{ij}(t)$ where $i$ is the index of the satellite hosting the optical bench, and $j$ is the index of the far spacecraft. 

\begin{figure}[!h]
    \centering
    \includegraphics[width=\columnwidth, trim={0.6cm, 1cm, 1.1cm, 0.3cm}, clip]{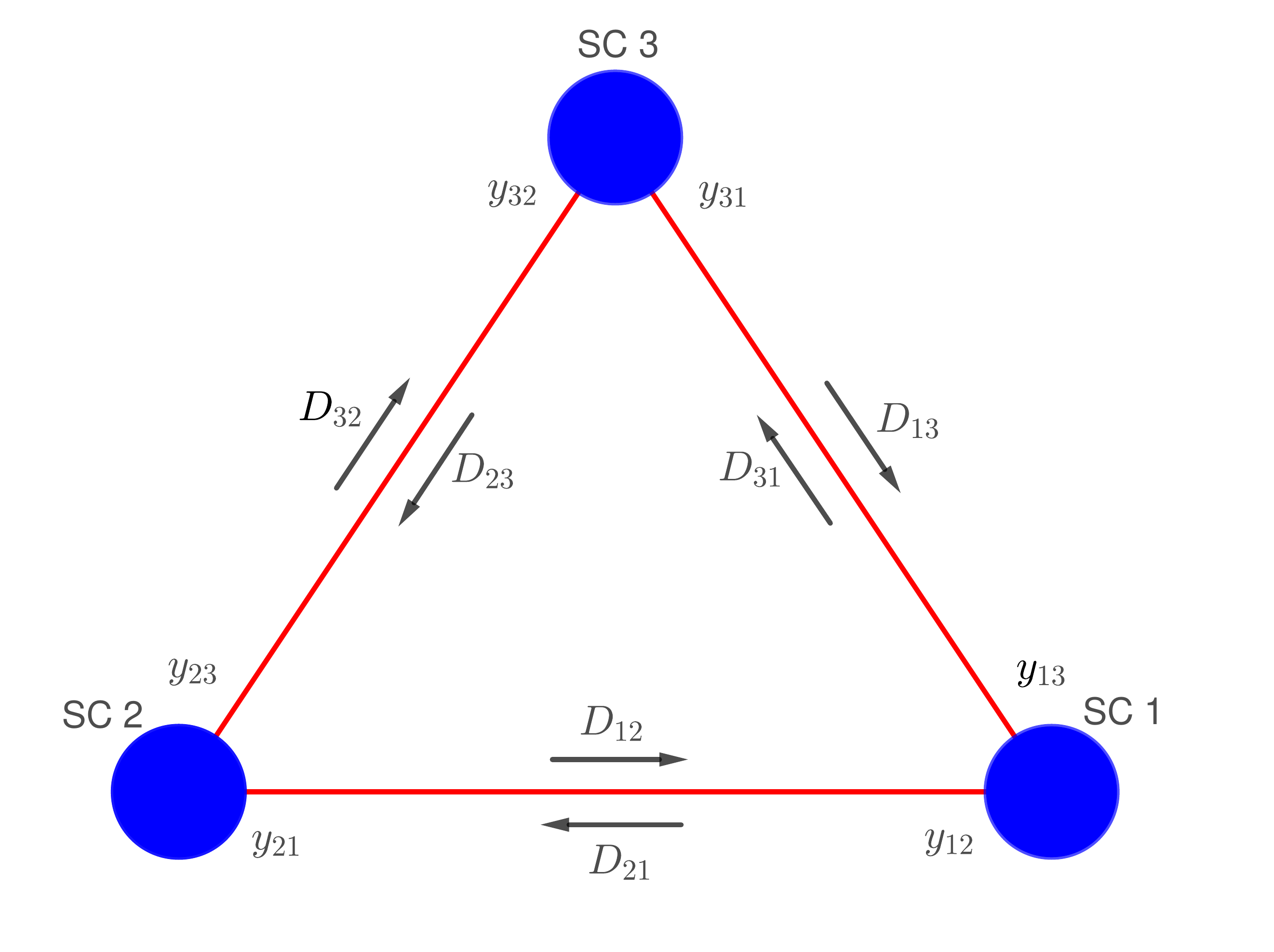}
    \caption{Schematic of the LISA constellation. The blue disks represent the three spacecrafts, the $y_{ij}$ indicate the interferometric measurement at optical bench $i$ hosted by spacecraft $i$ and receiving the laser beam from spacecraft $j$. The arrows correspond to the direction of the beams received by spacecraft $i$ and sent by spacecraft $j$, undergoing a time delay $D_{ij}$.}
    \label{fig:constellation}
\end{figure}

\subsection{\label{sec:tdi_filter}TDI as a filter}
Classical TDI algorithms usually operate in two steps. First, they compute delayed versions of discrete signals, interpolated at specific times depending on the light travel time delays along the constellation arms. In a second step, they combine these interpolated time series in such a way that laser noise terms vanish~\cite{tinto_cancellation_1999}.

Hence, each TDI channel $A_\alpha$ is produced by some \hl{linear combination of delayed versions of the data}:
\begin{equation}
{A}_{\alpha}(t)=\sum _{k=1}^{6}\sum _{n=1}^{{n}_{{\max},k}}{c}_{\textit{kn}\alpha}(t)\mathcal{D}[{d}_{\textit{kn}\alpha}(t)]{y}_{k}(t) 
\end{equation}
where we have introduced a delay operator $\mathcal{D}[d]$ which realizes a delay of the $y$ data stream by time $d$. For conciseness, we \hl{denoted} $y_{k}(t) = y_{i_k j_k}(t)$ the 6 interferometer measurements at optical bench $i_k$ of the beam coming from spacecraft $j_k$. For each channel $A_\alpha$ the model provides some set of coefficients $\{c_{k n \alpha}\}$ and a corresponding set of delays $\{d_{k n \alpha}(t)\}$. A delay $d_{k n \alpha}(t)$ includes the application of one or multiple light travel times $D_{ij}$ from spacecraft $j$ to spacecraft $i$ (see Fig.~\ref{fig:constellation} or Ref.~\cite{bayle_adapting_2021}), so that $\mathcal{D}[d_{k n \alpha}(t)] = \prod_{p} \mathcal{D}[{D}_{i_{knp} j_{knp}}(t)]$. In general, as the constellation evolves, these delays slowly vary with time~\cite{cornish_motion_2003}.

In practice, we do not have continuous data, but $N$ data points sampled at times $\{t_i\}$ with a sampling \hl{time}~$\tau_s$. We thus realize the delay operator by a fractional delay filter of some finite half-width~$n_{h}$~\cite{shaddock_postprocessed_2004}. In discrete form we write
\begin{eqnarray}
{{A}_{i\alpha}} = {\sum _{k=1}^{6}\sum _{n=1}^{{n}_{\max,k}}\sum _{l=-{n}_{\mathrm{h}}}^{{n}_{\mathrm{h}}}{c}_{\textit{kn}\alpha}({t}_{i}){f}_{\textit{lkn}\alpha}({t}_{i})\left[{D}_{l}{y}_{k}\right]({t}_{i})}
\label{eq:TDI_data_matrix} 
\end{eqnarray}
where the $f_{lkn\alpha}(t)$ are the fractional delay filter coefficients for the specified delay $d_k(t)$, and $D_l y_{k}$ is the data channel $k$ shifted by an integer number of samples $l$. 
To relate these combinations with the PCA formalism, we define the data matrix combining all shifted versions of the measurements $\bm{Y}$:
\begin{eqnarray}
\label{eq:x_matrix}
    \bm{X} \equiv \left({D}_{-{n_h}}\bm{Y},\,  \hdots, \, {D}_{+{n_h}}\bm{Y}\right),
\end{eqnarray}
where the $N \times M$ matrix $\bm{Y}$ gathers the $M$ measurements recorded at $N$ time samples. In what follows, we assume that it includes $M = 6$ measurements as
\begin{align}
	\label{eq:y_vector}
	\bm{Y} \equiv \left(\bm{y}_{12}, \, \bm{y}_{23}, \, \bm{y}_{31}, \, \bm{y}_{13}, \, \bm{y}_{21}, \, \bm{y}_{32} \right).
\end{align}
We also define the 6-row vectors $\bm{g}_{l \alpha}(t)$ with entries ${g}_{\textit{kl}\alpha}(t)\equiv \sum _{n=1}^{{n}_{\max,k}}{c}_{\textit{kn}\alpha}(t){f}_{\textit{lkn}\alpha}(t)$. For convenience, we gather them into a single-column vector as
\begin{equation}
    \bm{g}_{\alpha}(t) = \left(\bm{g}_{-{n_{h}}\alpha}(t) \, \hdots \, \bm{g}_{{+n_{h}}\alpha}(t)\right)^{T}.
\end{equation}
The size of matrix $\bm{X}$ is $N \times 6p$, where $p = 2n_h+1$ is the number of integer shifts, hence $\bm{g}_{\alpha}(t)$ has size $6p$.
With these definitions, Eq.~\eqref{eq:TDI_data_matrix} becomes
\begin{eqnarray}  
    A_{i\alpha}&=& \bm{X}_{i} \bm{g}_{\alpha}(t_i),
\label{eq:TDI_data_matrix_2}    
\end{eqnarray}
where we labelled by $\bm{X}_{i}$ the $\rm i^{th}$ row of matrix $\bm{X}$.

The aPCI treatment of Paper I introduced a data-driven approach to deriving specific alternatives to the $A_{i\alpha}$ by exploring more general linear combinations of the $X_{ij}$ which appear in Eq.~\eqref{eq:TDI_data_matrix_2}, seeking those combinations which minimize the sample variance and thus cancel the dominant noise. Importantly, in that approach those linear combinations arising from the usual TDI channel (any TDI generation in general) were a subspace of the possible broader space explored by the PCA treatment \emph{as long as} the coefficients $\bm{g}_{l\alpha}$ were effectively time-independent. 

For LISA, or similar instruments, the TDI coefficients do however slowly vary in time as the constellation flexes and evolves. This evolution limits the length of data matrix to which the analysis can be effectively applied, and thereby limits significantly the quality of the result. The issue is superficially similar to the hierarchy of various TDI ``generations'' with first generation TDI applying to a frozen constellation~\cite{giampieri_algorithms_1996, tinto_cancellation_1999}, and second generation accounting for leading-order temporal variation in the constellation~\cite{shaddock_motion_2003}. Our issue is distinct though. Any generation of TDI filters fits instantaneously into the form of the aPCI data matrix. The trouble here is that, at any order, those filters may be slowly time-dependent.  

The natural resolution is to allow similarly time-dependent linear combinations of the $\bm{X}_{ij}$ in the analysis. To see how to do this, consider sufficiently short segments where these evolving component values can be approximated linearly in time. We can Taylor expand the coefficents, writing $\bm{g}_{l\alpha}(t_i)\approx\bm{g}^{(0)}_{l\alpha}+(t_i-t_0)\bm{g}^{(1)}_{l\alpha} + \hdots$. Applying this in Eq.~\eqref{eq:TDI_data_matrix} we can then approximate the TDI construction as 
\begin{eqnarray}
    A_{i\alpha}&=& \sum_{q=0}^{m} \bm{X}_{i} \bm{g}^{(q)}_{l\alpha} (t_i-t_0)^q\nonumber\\
    &=& \bm{Z}_{i}^{(m)} \bm{G}_{\alpha}^{(m)},
\label{eq:TDI_data_matrix_Taylor}    
\end{eqnarray}
where we set $\bm G^{(m)} \equiv \left({\bm{g}^{(0)\dag}},\,  {\bm{g}^{(1)\dag}},\, \hdots , \, {\bm{g}^{(m)\dag}} \right)^{\dag}$ and the new data matrix of order $m$ whose rows are given by
\begin{equation}
    \label{eq:z_matrix}
    \bm{Z}_{i}^{(m)} = \left(\bm{X}_{i}, \, (t_i-t_0)\bm{X}_{i}, \,  \hdots,  \, (t_i-t_0)^{m}\bm{X}_{i} \right).
\end{equation}
With this convention, we have $\bm{Z}^{(0)} = \bm{X}$. The representation now has the same form as in Eq.~\eqref{eq:TDI_data_matrix}, but with time-independent coefficients, and with a slightly more complicated data matrix $\bm Z^{(m)}$ which includes a copy of $\bm X$ together with essentially $t^{q}\times\bm X$, with $q$ running from 1 to $m$. Then we can proceed with the PCA analysis as in Paper I but with a data matrix that is now enlarged by a factor of $2m$ on its short side. 
Although Eq.~\eqref{eq:z_matrix} describes an arbitrary Taylor expansion, we will be working with first order in time ($m=1$) in the following.

\subsection{\label{sec:pca_time}Principal component analysis of the data matrix}

TDI \hl{combinations} are derived from the knowledge of the design matrix, i.e., the way laser noise sources enter the interferometric data, including the exact delays and the time variations they undergo. In other words, TDI fully specifies the matrix of filter coefficients $\bm{G}_{\alpha}$ introduced in Eq.~\eqref{eq:TDI_data_matrix_Taylor}. In the aPCI approach, the noise-cancelling decomposition is derived from the data. To do so, we proceed as in Paper I with $\bm Z^{(m)}$ instead of $\bm X$, so that the higher-order term (or terms) in time are now present in the data matrix. Then, we compute its singular value decomposition (SVD):
\begin{eqnarray}
\label{eq:z_svd}
    \bm{Z}^{(m)} = \bm{U}^{(m)} \bm{S}^{(m)} \bm{V}^{(m)\dag},
\end{eqnarray}
where $\bm{U}^{(m)}$ and $\bm{V}^{(m)\dag}$ are unitary matrices whose columns are basis vectors we will refer to as singular vectors, and \hl{$\bm{S}^{(m)}$} is a $N \times 6p(m+1)$ rectangular diagonal matrix whose elements are positive real numbers called singular values. 
We obtain the principal components (PCs) of $\bm{Z}^{(m)}$ by applying the transformation
\begin{equation}
\label{eq:e_matrix}
    \bm{E}^{(m)} = \bm{Z}^{(m)} \bm{V}^{(m)}.
\end{equation}
The columns of $\bm{E}^{(m)}$ form the aPCI combinations, and are ordered from the lowest to the largest singular value associated to them. This ordering of PCA components is in reverse order from typical convention, as most PCA applications value information in the high-variance components. Note that this convention was not adopted in the beginning of Paper I, where the transformed data matrix was labelled $\bm{T}$ instead of $\bm{E}$. Singular values are a measure of the variance carried by each component. Selecting the lowest-variance components, therefore, provides us with combinations where the laser frequency noise is minimal. In the following, $\bm{e}_{j}^{(m)}$ denotes the $j^{\rm th}$ lowest variance aPCI variable, such that its entries are $e_{j}^{(m)}(n) = E\left(n, j\right)$.

\section{\label{sec:laser_noise_mitigation}Laser noise mitigation}

To evaluate the performance of the first-order aPCI, we need to i) verify the proper mitigation of laser noise by the combinations $\bm{e}_{j}^{(1)}$ and ii) check that their sensitivity to GWs is comparable to second-generation TDI's, which is the state-of-the-art technique to cancel laser frequency noise terms for a flexing constellation, up to first order in time delay derivatives. In this section, we assess the level of laser noise cancellation.

\subsection{\label{sec:noise_simulation}Data simulation}

We use a simulation of the 18 interferometric outputs measured by LISA while in orbit, assuming a non-equal arm, flexing constellation following Keplerian orbits around the Sun. These outputs include measurements from the science, reference, and test-mass interferometers, as defined, e.g., in Ref.~\cite{bayle_effect_2019}. Furthermore, we assume that the six lasers are independent, which means that laser locking is off. We perform this simulation with Bayle et al.'s \textsc{LISA Instrument} simulator~\cite{lisainstrument}, a Python-based simulator cross-checked against the \textsc{LISANode} simulator~\cite{bayle:tel-03120731}. Only laser noise is present in the simulation, with an amplitude spectral density of 28.2 $\rm Hz.Hz^{-1/2}$ (we assume it is white in the simulation bandwidth). We set the sampling frequency of the output measurements to 4 Hz in accordance with LISA Science Requirements Document~\cite{lisa_simulation_working_group_lisa_2018}. The simulation runs at a cadence four times faster than the output sampling, and anti-aliasing filters are adjusted accordingly.

Instead of directly analyzing the 18 interferometer outputs, \hl{we reduce the problem's dimension} by condensing them into the 6 intermediary variables $\eta_{ij}$ using Staab et al.'s Python-based TDI calculator \textsc{pyTDI}\cite{pytdi}. This operation amounts to assuming a configuration with only 3 independent running lasers (see, e.g., \cite{otto_time-delay_2015} for a detailed definition). So we take $y_{ij} = \eta_{ij}$ in Eq.~\eqref{eq:y_vector}. 
\hl{In a future work, we plan to address the more realistic configuration with 6 locked lasers, by analyzing the full set of interferimeter data through aPCI.}

We generate the secondary noises (i.e., non-laser-frequency noises) independently for the 6 intermediary variables and add them to the simulation outputs, \hl{ignoring any correlation among them}. While this operation does not realistically reflect how secondary noises propagate through the instrument, it allows us to perfectly control the content of the simulation. \hl{Furthermore, these assumptions only affect secondary noise, whereas our study focuses on reducing laser frequency noise. Thus we do not expect them to impact the results presented here, especially since the secondary noise PSDs are assumed to be known in the following.}
In our setup, we assume the presence of two secondary noises: a noise due to the residual test-masses' (TMs) accelerations with respect to the inertial frame, of PSD $S_{\mathrm{a}}(f)$, and noises coming from the residual displacement in the optical metrology system (OMS), including position readouts. Hence, we can write the secondary noise PSDs as 
\begin{equation}
\label{eq:secondary_noises_psd}
    S_{n}(f) = S_{\mathrm{OMS}}(f) + 2 S_{\mathrm{a}}(f),
\end{equation}
where the factor of 2 comes from the fact that the test-mass noise appears in the $\eta_{ij}$ from both the contribution of link $ij$ via the science interferometer and from link $ji$ via the test-mass interferometer. We provide the analytical expressions for the acceleration and OMS noise PSDs in Appendix~\ref{sec:secondary_noises}.

\subsection{\label{sec:noise_mitigation_analysis}Analysis of simulated data}

Once the noisy variables $\eta_{ij}$ are generated, we form the data matrix $\bm{Z}^{(m)}$ with $m = \{0, 1\}$ as defined by Eqs.~\eqref{eq:x_matrix} and \eqref{eq:z_matrix}. We consider 12 hours worth of data, which is long enough to allow for a significant variation of the armlengths and probe the lowest frequencies of LISA's bandwidth. We choose a half-width (or stencil size) large enough to encompass the number of time delays applied in second-generation TDI, to which we add a margin corresponding to the order of Lagrange polynomials typically used in TDI. This translates as $n_h = \lfloor 8 L / (c \tau_s) \rfloor + 32$. With $L = 2.5$ Gm as the average arm length, $\tau_s = 0.25$ s the sampling cadence and $c$ is the speed of light, we get $n_h = 266$.

Then, we compute the PCA of $\bm{Z}^{(m)}$ as described in Eq.~\eqref{eq:z_svd} using the \textsc{scikit-learn} Python package~\cite{pedregosa_scikit-learn_2011}, which runs the full SVD with the standard LAPACK solver. This package also features incremental principal components analysis (IPCA) which allows us to split the computation in chunks and optimize the memory usage.

\begin{figure}[!h]
    \centering
    \includegraphics[width=\columnwidth, trim={0.5cm, 0.3cm, 0.5cm, 0.3cm}, clip]{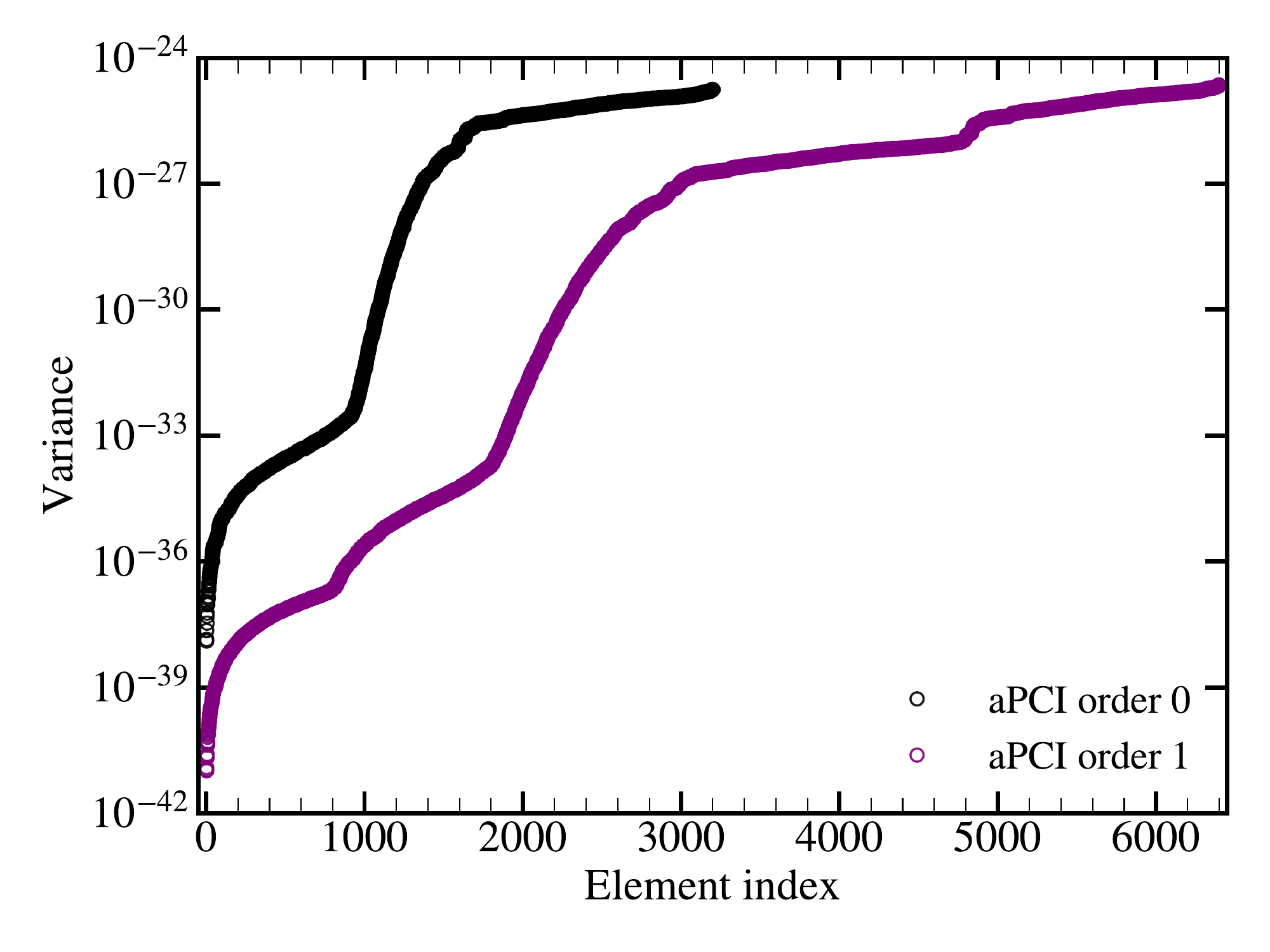}
    \caption{Normalized singular values as a function of the principal components conveying the amount of variance they carry. We plot the zeroth order (black) and first order (purple) cases. We observe a dynamic range between the highest and the lowest variances that is more pronounced for the first-order decomposition, suggesting a better noise decomposition with first order than with zeroth order.
    }
    \label{fig:variances}
\end{figure}

We plot the \hl{amount of variance in each component (i.e. the singular values squared} divided by the number of data points $N$) for orders $m = 0$ and $m = 1$ in Fig.~\ref{fig:variances}. The PCs with the lowest variances should be the ones that best reject laser frequency noise. Note that we ordered them by increasing singular values here, that is why it appears flipped with respect to Fig.~1 of Paper~I. The zeroth order (represented in black) corresponds to the case studied in Paper~I, where no time variations are accounted for in the analysis (although they are present in the simulated data). The first-order curve (in purple) includes the extension to linear variations in time, and features a larger difference between the largest and the lowest variances. This indicates a more faithful decomposition of the noise \hl{into large variance and low variance components}. 

\begin{figure}[!h]
    \centering
    \includegraphics[width=\columnwidth, trim={0.5cm, 0.3cm, 0.5cm, 0.3cm}, clip]{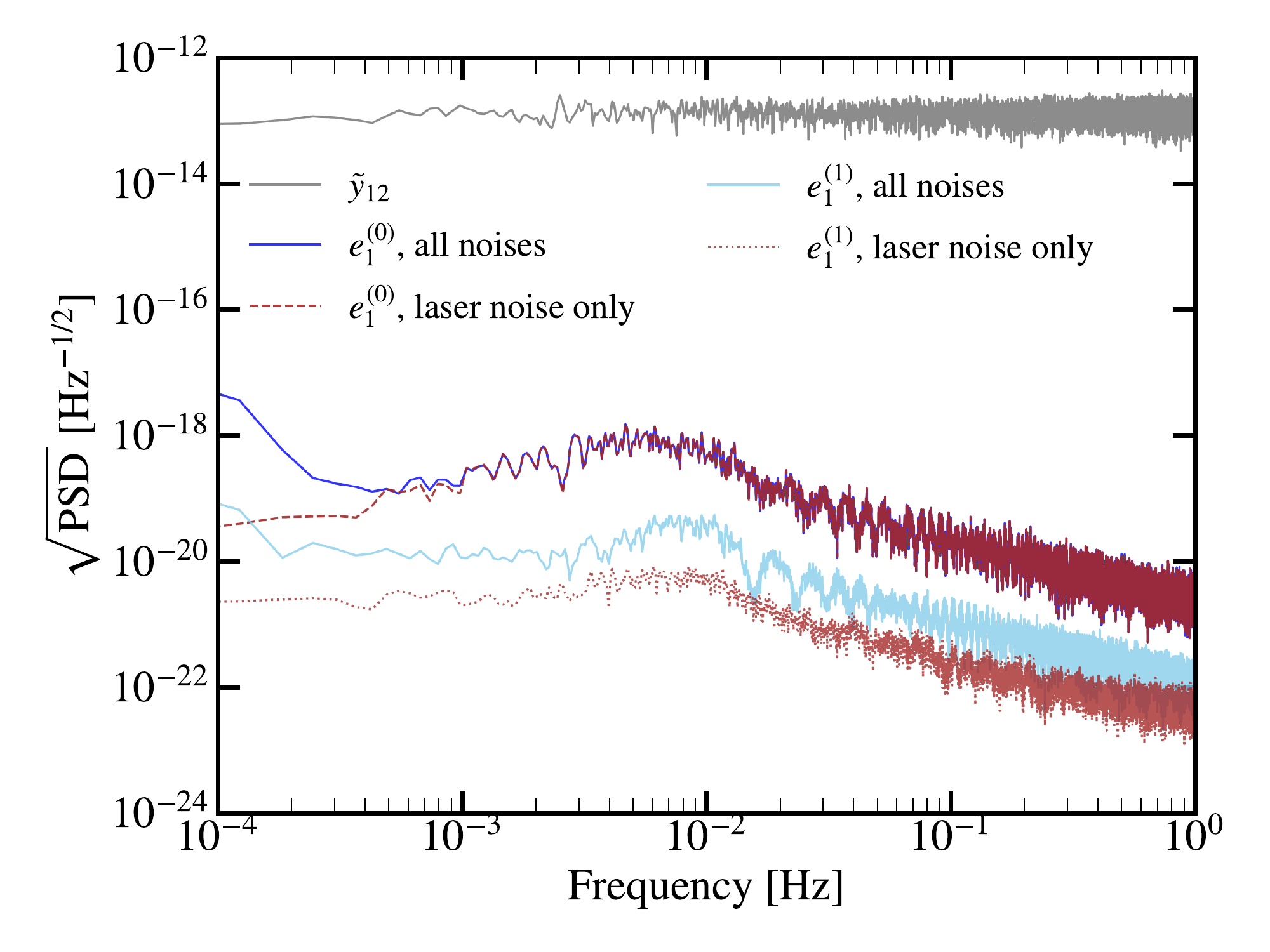}
    \caption{Periodogram of the lowest variance aPCI variable \hl{$\bm{e}_{j}^{(m)}$} for zeroth (dark blue) and first (light blue) order decompositions, along with channel $12$ of the input data vector $\bm{y}$. The red dotted curves show the contribution of laser noise for both orders. The laser noise's contribution to the residuals in the first-order case is lower than the all-noise residuals, showing that laser noise is suppressed below other noises.}
    \label{fig:aPCI_noise_periodograms}
\end{figure}

For gravitational-wave detection, we only need the PCs with lowest singular values. As in Paper~I, we select the q lowest variance components beyond there is no meaningful improvement in GW sensitivity, and then project the data onto these components. As an example, we plot the periodogram of the aPCI variable $\bm{e}_{1}^{(m)}$ (the one with lowest variance) in Fig~\ref{fig:aPCI_noise_periodograms} for $m = 0$ (dark blue) and $m = 1$ (light blue). We also plot the periodogram of the single-link measurement $\bm{y}_{12}$ (gray).

We observe a difference of 7 to 8 orders of magnitude between the input data and the first-order aPCI variable, while the noise level of the zeroth order variable is about ten times larger than its first-order counterpart. This result suggests that including the first-order terms in the aPCI process helps to cancel laser noise with time-varying armlengths. To confirm, we apply the same transformation to simulated data containing only laser frequency noise (with no secondary noise contribution). The result is shown by the light dotted and dark dashed red lines Fig~\ref{fig:aPCI_noise_periodograms} corresponding to zeroth and first order, respectively. In the case of order zero, the aPCI projection of the laser-frequency-noise-only data is almost superimposed on the projection of both laser and secondary noises, which confirms that laser frequency noise still dominates. On the contrary, the first-order case exhibits lower laser frequency noise residuals (light dotted red) compared to the data containing all noises (light solid blue). This result shows that the first-order extension of aPCI effectively \hl{reduces} laser frequency noise below the level of other noises.

\section{\label{sec:sensitivity}Sensitivity of first-order PCI combinations}

Suppression of laser-noise would not be helpful if it inadvertently also suppressed gravitational wave signals. In this section, we compute the GW sensitivity of aPCI combinations $\bm{E}^{(1)}$, so that we can compare the algorithm performance with standard TDI.

\subsection{\label{sec:noise_response}PCI response to stochastic processes}

 We analyze how a stochastic process measured in $\bm{y}_{ij}$ is transformed by the PCI combinations. Let us consider a zero-mean Gaussian, multivariate stationary process $\bm{Y}$ with 6 channels and  $N$ points in time. Taking the Fourier transform of $\bm{Y}$ allows us to work with covariances defined for each frequency bin $f$, neglecting the correlations between two different frequency bins. We thus consider the $6$-column vector $\bm{\tilde{y}}$ defined as
\begin{equation}
\bm{\tilde{y}}(f) = \left(\tilde{y}_{12}, \, \tilde{y}_{23}, \, \tilde{y}_{31}, \, \tilde{y}_{13}, \, \tilde{y}_{21}, \, \tilde{y}_{32} \right)^T,
\end{equation}
where the convention \hl{$\bm{\tilde{y}}(f)$} refers to the DFT of any vector $\bm{y}$ at frequency $f$.

As in Paper~I, we can write transformation in Eq.~\eqref{eq:e_matrix} in the Fourier domain at zeroth order in time. For any frequency $f$, we relate the $6p$-vector of aPCI variables $\bm{\tilde{e}}(f)$ to the $6$-vector $\bm{\tilde{y}}(f)$ through the simple matrix operation
\begin{equation}
\label{eq:e_response}
    \bm{\tilde{e}}^{(0)}(f) = \bm{\tilde{W}}(f) \bm{\tilde{y}}(f),
\end{equation}
where we defined the $6p \times 6$ transformation matrix $\bm{\tilde{W}}(f)$
\begin{equation}
\label{eq:x_vector}
    \bm{\tilde{W}}(f) = \bm{V}^{\dag} \bm{\tilde{\Omega}}(f),
\end{equation}
where $\bm{\tilde{\Omega}} \equiv \begin{pmatrix} \bm{\tilde{\Omega}}_{-n_h} & \hdots & \bm{\tilde{\Omega}}_{+n_h}  \end{pmatrix}^T$ is a $6p \times 6$ matrix including the $6 \times 6$ blocks $\bm{\tilde{\Omega}}_{l}(f)$ which encode the application of delay $l$ to all channels. Each $\bm{\tilde{\Omega}}_{l}$ is a  diagonal matrix constructed as
\begin{equation}
    \bm{\tilde{\Omega}}_{l} = \mathrm{diag}\left(\tilde{D}_{l}, \, \hdots, \, \tilde{D}_{l} \right),
\end{equation}
whose diagonal elements are the Fourier-domain approximation of the delay operators:
\begin{equation}
\label{eq:delay_operator}
    \tilde{D}_{l}(f) = e^{-2\mathrm{i} \pi f l \tau_s},
\end{equation}
\hl{where we labeled the complex number as $\mathrm{i} = \sqrt{-1}$.}
If $\bm{y}$ is a stationary stochastic process of covariance $\bm{\tilde{\Sigma}}_{y}(f)$, then the covariance matrix of $\bm{\tilde{e}}$ is
\begin{eqnarray}
\label{eq:e_covariance}
    \bm{\tilde{\Sigma}}_{e}(f)  \approx \bm{\tilde{W}}(f) \bm{\tilde{\Sigma}}_{y}(f) \bm{\tilde{W}}(f)^{\dag}.
\end{eqnarray}
Rigorously, we should then account for the first-order part of the data matrix $\bm{Z}^{(1)}$, as given in Eq.~\eqref{eq:z_matrix}. However, while this part is obviously important to mitigate the linear variations of the laser frequency noises, its effect is not dominant when considering the response to secondary noises and gravitational waves (which are several orders of magnitude smaller than laser noise). Hence, in the following we neglect the contribution of time variations when computing the covariance of the aPCI variables. We will discuss in Section~\ref{sec:sensitivity_results} the consequences of this simplification.

With Eq.~\eqref{eq:e_response}, we established the recipe to propagate any GW waveform from its single-link responses to the aPCI variables. Likewise, we can use Eq.~\eqref{eq:e_covariance} to convert the spectrum of any secondary noise (that is not laser frequency noise) into its aPCI spectrum. 

\subsection{\label{sec:laser_projection}Laser frequency noise projection}

Laser frequency noise, as all other components in the data, projects onto the basis of singular vectors through Eq.~\eqref{eq:e_matrix}. We can mitigate its impact in any further analysis by simply considering only the first $q$ lowest singular value components, and discarding all the others. This is the counterpart of what is commonly called ``truncated PCA", which usually discards the lowest singular values. In Paper~I, we determined that the aPCI sensitivity was increasing until $q = 6$. Including additional components did not improve it further, as laser frequency noise starts to dominate at larger $q$. We adopt this cut-off in the following.

\subsection{\label{sec:orthogonalization}Orthogonalization}

We use Eq.~\eqref{eq:e_covariance} to compute the $q \times q$ covariance $\bm{\tilde{\Sigma}}_{{e}_{n}}(f)$ of the q lowest-variance aPCI variables.
We assume that all single-link noises are uncorrelated and have the same PSD, set by Eq.~\eqref{eq:secondary_noises_psd}. Hence, their covariance $\bm{\tilde{\Sigma}}_{y}(f)$ is diagonal. In the same way as for Michelson TDI, the aPCI transformation introduces correlations among the resulting  variables, so that the matrix $\bm{\tilde{\Sigma}}_{{e}_{n}}(f)$ has non-zero off-diagonal terms.
Exactly like when we construct TDI A, E, T, another transformation is needed if we want to work with orthogonal data streams. One can perform this transformation by decomposing  $\bm{\tilde{\Sigma}}_{{e}_{n}}(f)$ into its eigenbasis. It turns out that $\bm{\tilde{\Sigma}}_{{e}_{n}}(f)$ has only three non-zero eigenvalues. This is a consequence of the secondary noise approximation we made in Eq.~\eqref{eq:e_covariance}, which neglects laser frequency noise residuals.  \hl{Without this assumption, the covariance has actually three other non-null eigenvalues, which are however much smaller than the first three. Indeed, the algorithm learns the correlations in the data up to a statistical error, which is reflected in an unperfect separation of laser frequency noise and other noises. However, as we shall see, the eigenstreams corresponding to the largest eigenvalues will be sufficient for most analyses, as they carry the bulk of the sensitivity.} We define the eigenstreams as the projection of the initial variables onto the eigenspace via
\begin{equation}
\label{eq:orthogonalization}
    \bm{\tilde{e}}^{(m)}_{\perp}(f) = \bm{\Phi}^{\dag}(f) \bm{\tilde{e}}^{(m)}(f),
\end{equation}
where $\bm{\Phi}(f)$ is the matrix of the covariance's eigenvectors $\bm{\tilde{\Sigma}}_{{e}_{n}}(f)$:
\begin{equation}
\label{eq:covariance_eigendecomposition}
\bm{\tilde{\Sigma}}_{{e}_{n}}(f) = \bm{\Phi}(f) \bm{\Lambda}(f) \bm{\Phi}(f)^{\dag},
\end{equation}
with $\bm{\Lambda}(f)$ being the diagonal matrix of eigenvalues, which are proportional to the PSDs of the orthogonal variables:  $\Lambda_{lp}(f) = \langle | \, \tilde{e}^{(m)}_{\perp l}(f) |^{2} \rangle \delta_{lp}$.

\begin{figure}[!h]
    \centering
    \includegraphics[width=\columnwidth, trim={0.5cm, 0.3cm, 0.5cm, 0.3cm}, clip]{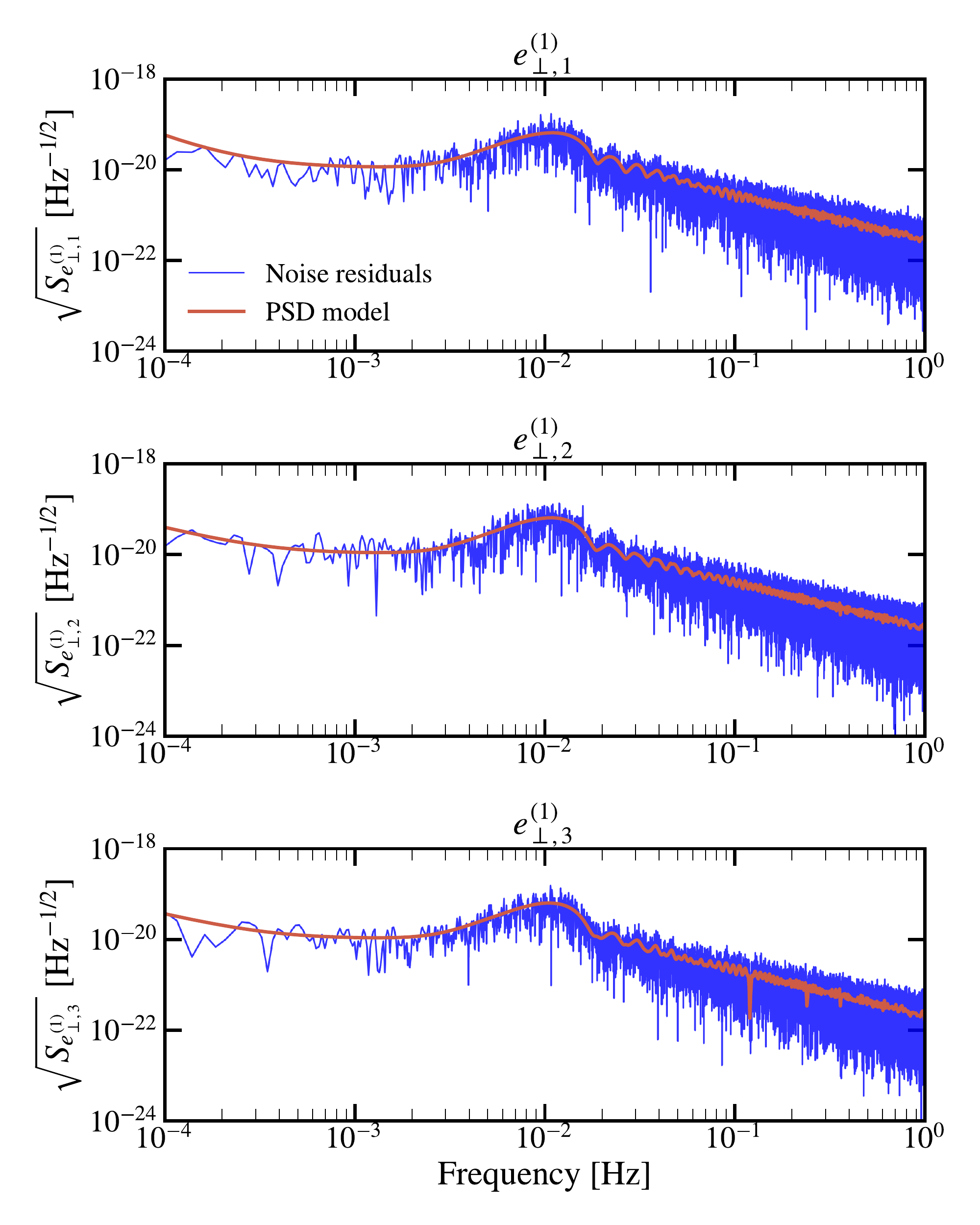}
    \caption{Noise periodograms of the 3 orthogonalized first-order aPCI variable $\bm{e}_{\perp j}^{(1)}$ (in blue) along with their analytical power spectra (in red).}
    \label{fig:aPCI_noise_models}
\end{figure}

To verify our modeling, we plot (in blue) the periodograms of the three aPCI eigenstreams associated with non-zero eigenvalues in Fig.~\ref{fig:aPCI_noise_models}. 
We compute their analytical covariance using the zeroth-order approximation in Eq.~\eqref{eq:e_covariance} and plot its non-zero eigenvalues in red. The model overlaps well with the periodograms, hereby showing that it is acceptable to only account for zeroth-order effects when computing residual noise aPCI spectra.

\subsection{\label{sec:gw_response}PCI response to gravitational waves}

Up to this point, no information about the constellation orbits has been required, since we derived the transformation matrix $\bm{\tilde{W}}$ directly from the data. That said, the link response to a GW point source with propagation vector $\bm{k}$ and Fourier amplitudes $\tilde{h}_{+}(f', \bm{k}), \tilde{h}_{\times}(f', \bm{k})$ depends on the spacecrafts orbits, as shown in Eq.~(29) of Paper I, which we reproduce here:
\begin{align}
	\label{eq:arm_esponse_frequency}
	y^{\mathrm{GW}}_{ij}(t, \bm{k}) & = \sum_{\alpha=+,\times} \int_{-\infty}^{+\infty}  \tilde{h}_{\alpha}\left(f', \bm{k}\right) e^{2\mathrm{i} \pi f' t} \frac{  F_{\alpha}\left(  \psi , \bm{k}, \bm{n}_{ij}\right)}{2\left(1 - \bm{k}\cdot \bm{n}_{ij}\right))}  \nonumber \\
	& \times \left[ e^{- 2\mathrm{i} \pi f' \left(L_{ij} + \bm{k} \cdot \bm{r}_{j}(t_j)\right) / c}- e^{- 2\mathrm{i} \pi f' \bm{k} \cdot \bm{r}_{i}(t)/ c} \right] df'.
\end{align}
The orbits are needed to determine the spacecraft position vectors $\bm{r}_{i}(t)$ and the constellation arms orientation vectors $\bm{\hat{n}}_{ij} = \left(\bm{r}_{i} - \bm{r}_{j}\right) / \lVert \bm{r}_{i}- \bm{r}_{j} \rVert$ as a function of time. However, the response requires far less precise information than TDI: while a nanosecond accuracy is needed for TDI, a 100 ms accuracy is likely to be sufficient for most GW sources~\cite{katz2022}. The response also depends on the sky location through $\bm{k}$ and the polarization angle $\psi$. For any time $t$, we can conveniently express the equation above as a matrix relation, 
\begin{equation}
\label{eq:y_gw_matrix}
\bm{\tilde{y}}_{\mathrm{GW}}(t, \bm{k}) = \int_{-\infty}^{+\infty} \bm{\tilde{H}}(t, f', \bm{k}) \bm{\tilde{h}}(f', \bm{k}) df',
\end{equation}
where $\bm{\tilde{H}}(f, \bm{k})$ is the $6 \times 2$ GW response matrix with as many rows \hl{as} there are links, and as many columns as there are polarization modes: 
\begin{align}
	\label{eq:h_matrix}
\tilde{H}_{n, p}(t, f', \bm{k}) & = e^{2\mathrm{i} \pi f' t} \frac{  F_{\alpha(p)}\left(  \psi , \bm{k}, \bm{n}_{i(n)j(n)}\right)}{2\left(1 - \bm{k}\cdot \bm{n}_{i(n)j(n)}\right))}  \nonumber \\
& \times \left[ e^{- 2\mathrm{i} \pi f' \left(L_{i(n)j(n)} + \bm{k} \cdot \bm{r}_{j(n)}(t_{j(n)})\right) / c} \right.
\nonumber \\ 
& \left. - e^{- 2\mathrm{i} \pi f' \bm{k} \cdot \bm{r}_{i(n)}(t)/ c} \right].
\end{align}
\hl{The rows of the matrix are indexed by $n$ so that $i(n)j(n)$ follows the same ordering as the entries of vector $\bm{\tilde}{y}(f)$ in} Eq.~\refeq{eq:y_vector}. \hl{The columns are indexed by $p$ such that $\alpha(1) = +$ and $\alpha(2) = \times$.} We also defined the vector of strain amplitudes \hl{$\bm{\tilde{h}}(f, \bm{k}) \equiv \big(\tilde{h}_{+}(f, \bm{k}), \tilde{h}_{\times}(f, \bm{k})\big)^{\intercal}$}.

To easily compare the analytical response with simulated data, we study the case of an isotropic, stationary, zero-mean Gaussian stochastic GW background with a strain PSD equal to unity. The power spectrum of the strain amplitudes is then~\cite{caprini_reconstructing_2019}
\begin{align}
\label{eq:h_covariance}
   \langle \tilde{h}_{\alpha}(f, \bm{k}) \tilde{h}_{\alpha}(f', \bm{k}')^{\ast} \rangle & = \nonumber \\
   \frac{1}{2} \delta\left(f - f'\right) \frac{1}{4\pi} \delta\left(\bm{k} - \bm{k}'\right) \delta_{\alpha, \alpha'} S_{\mathrm{GW}}(f),
\end{align}
where we defined the power spectrum operator $\langle \cdot \rangle$ in Appendix~\ref{eq:average_operator}. 
We are interested in the sky-averaged response $\bm{\tilde{y}}_{\mathrm{GW}} = \int_{\bm{k}}\bm{\tilde{y}}_{\mathrm{GW}}(\bm{k}) d^{2} \bm{k}$. Using Eqs.~\eqref{eq:y_gw_matrix} and~\eqref{eq:h_covariance}, we can then write the sky-averaged link response $\bm{R}_{y}(f)$ as
\begin{align}
\label{eq:averaged_y_gw_response}
\bm{R}_{y}(f) & \equiv \langle\bm{\tilde{y}}_{\mathrm{GW}}(f) \bm{\tilde{y}}_{\mathrm{GW}}(f)^{\dag} \rangle \nonumber \\
& = \frac{1}{8\pi} \int_{\bm{k}} \bm{\tilde{H}}(f, \bm{k}) \bm{\tilde{H}}(f, \bm{k})^{\dag} d^{2} \bm{k}.
\end{align}

Therefore, the link response to the GW background is a stochastic process of covariance given by Eq.~\ref{eq:averaged_y_gw_response}. To compute the aPCI GW response, we simply need to use Eq.~\eqref{eq:e_covariance}, which yields 
\begin{eqnarray}
\label{eq:e_gw_response}
    \bm{R}_{e}(f)  = \bm{\tilde{W}}(f) \bm{R}_{y}(f) \bm{\tilde{W}}^{\dag}(f).
\end{eqnarray}
Similarly, it follows from Eq.~\eqref{eq:orthogonalization} that the GW response from orthogonalized aPCI variables is 
\begin{eqnarray}
\label{eq:e_gw_ortho_response}
    \bm{R}_{e, \perp}(f)  = \bm{\Phi}^{\dag}(f) \bm{R}_{e}(f) \bm{\Phi}(f).
\end{eqnarray}

\begin{figure}[!h]
    \centering
    \includegraphics[width=\columnwidth, trim={0.5cm, 0.3cm, 0.5cm, 0.3cm}, clip]{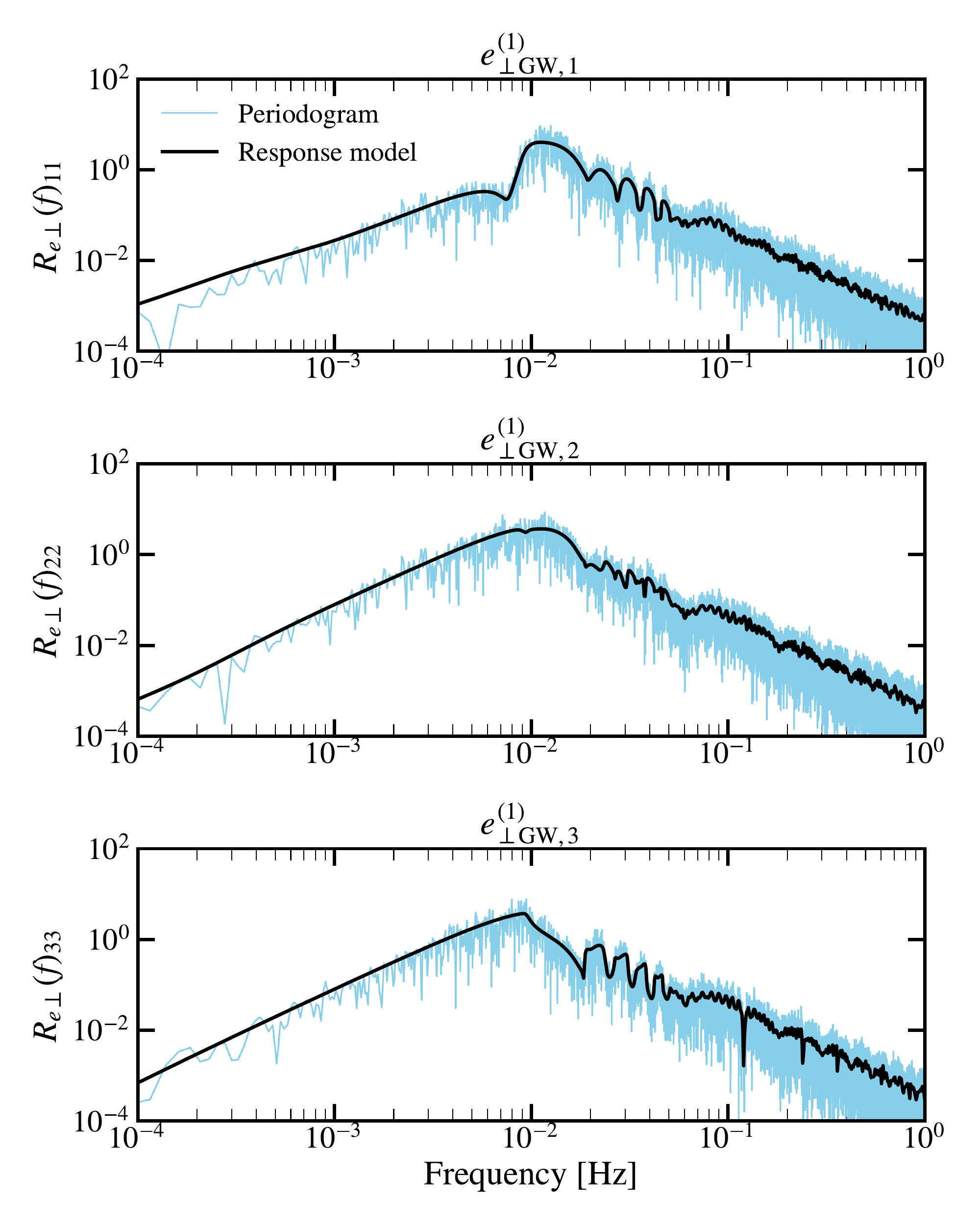}
    \caption{GW response periodograms of the 3 first-order orthogonalized aPCI variables $\bm{e}_{\perp \mathrm{GW}, j}^{(1)}$ to a stochastic GW background with a strain PSD of $\rm 1 \, Hz^{-1}$ (blue), along with their analytical response function (black).}
    \label{fig:aPCI_gw_response}
\end{figure}

To check the validity of the response function, we simulate an isotropic stochastic gravitational-wave background using \texttt{LISA GW Response}~\cite{lisagwresponse}, assuming independent polarizations and a PSD equal to unity, as described by Eq.~\eqref{eq:h_covariance}. In the simulation, the background stems from 768 independent point-sources dividing the sky into \hl{the same number of} pixels, distributed on a \texttt{HEALPix} map~\cite{Gorski_2005}. Then, we apply the exact same aPCI transformation that we obtained in Sec.~\ref{sec:noise_mitigation_analysis} to the measured link responses $\bm{\tilde{y}}_{\mathrm{GW}}$. We get the GW signal as seen through the aPCI variables $\bm{\tilde{e}}_{\mathrm{GW}}$, which we project onto the eigenspace of the secondary noise covariance using Eq.~\eqref{eq:orthogonalization}. We finally obtain the frequency series $\bm{\tilde{y}}_{\perp, \mathrm{GW}}(f)$. We plot their periodogram in light blue in Fig~\ref{fig:aPCI_gw_response} and we superimpose the analytical response that we derived with Eq.~\ref{eq:e_gw_ortho_response}.

 We check that the model matches the simulation by inspecting the distributions of the real and imaginary parts of the Fourier transforms normalized by their response $2 \bm{R}_{e, \perp}^{-1} \bm{\tilde{y}}_{\perp, \mathrm{GW}}$. We verify that they follow a normal distribution of mean zero and unit standard deviation. Like for the noise residuals, this result shows that the zeroth-order analytical model is sufficient to describe the aPCI response to GWs.

\subsection{\label{sec:sensitivity_definition}Computation of sensitivity}
In the GW literature, sensitivity is commonly defined as the ratio of the PSD of the noise affecting the measurement to the instrument's sky-averaged response function. Here we strictly follow the definition of Babak et al.~\cite{babak_lisa_2021}. The sensitivity of one variable $\bm{e}_{\perp j}^{(1)}$ is therefore 
\begin{equation}
\label{eq:t_sensitivity}
    {S}_{h, e_{\perp j}}(f) = \frac{\Lambda_{jj}(f)}{R_{e, \perp jj}(f)},
\end{equation}
where the numerator is given by the aPCI covariance eigenvalues in Eq.~\eqref{eq:covariance_eigendecomposition}, and the denominator is the diagonal element of the response in Eq.~\eqref{eq:e_gw_ortho_response}. Making use of orthogonality, we derive the total sensitivity by summing the inverse sensitivities (i.e., the signal-to-noise ratios or SNRs) of the individual variables as
\begin{equation}
\label{eq:total_sensitivity}
    {S}_{h, e_{\perp}}(f) = \left[ \sum_{j=1}^{3}  {S}^{-1}_{h, e_{\perp j}}(f)  \right]^{-1}.
\end{equation}

In the following, we will compare the aPCI sensitivity that we obtain with standard second-generation TDI. To this aim, we compute the TDI sensitivity in a similar way. Instead of using the aPCI transformation matrix $\bm{\tilde{W}}(f)$, we build a TDI transformation matrix $\bm{\tilde{W}}_{\mathrm{TDI}}(f)$ in the frequency domain. The TDI-equivalent output of Eq.~\eqref{eq:e_response} is then a 3-vector whose elements are the Michelson variables $X$, $Y$ and $Z$. At zeroth order, we can derive the entries of $\bm{\tilde{W}}_{\mathrm{TDI}}(f)$ for the second generation TDI from Eq.~(106) in Babak et al.~\cite{babak_lisa_2021}, where we approximate all delays operators by their frequency-domain limit for infinite time series: $\tilde{D}_{ij} = e^{-2\pi i f L_{ij} / c}$. We evaluate the armlengths at half the observation time $L_{ij} = L_{ij}(N \tau_s / 2)$. Then, we compute the Michelson TDI covariance exactly as in Eq.~\eqref{eq:e_covariance} and we diagonalize it to obtain the orthogonal TDI variables. Performing this exact orthogonalization instead of using the standard A, E, T formula (as derived in~\cite{prince_lisa_2002} under specific hypotheses) ensures that we follow the same process for both aPCI and TDI to compute the sensitivity. Moreover, the standard definition does not yield perfectly orthogonal combinations in the case of non-equal armlengths~\cite{baghi_statistical_2021}.

\subsection{\label{sec:sensitivity_results}Sensitivity results}

\begin{figure}[!h]
    \centering
    \includegraphics[width=\columnwidth, trim={0.5cm, 0.3cm, 0.5cm, 0.3cm}, clip]{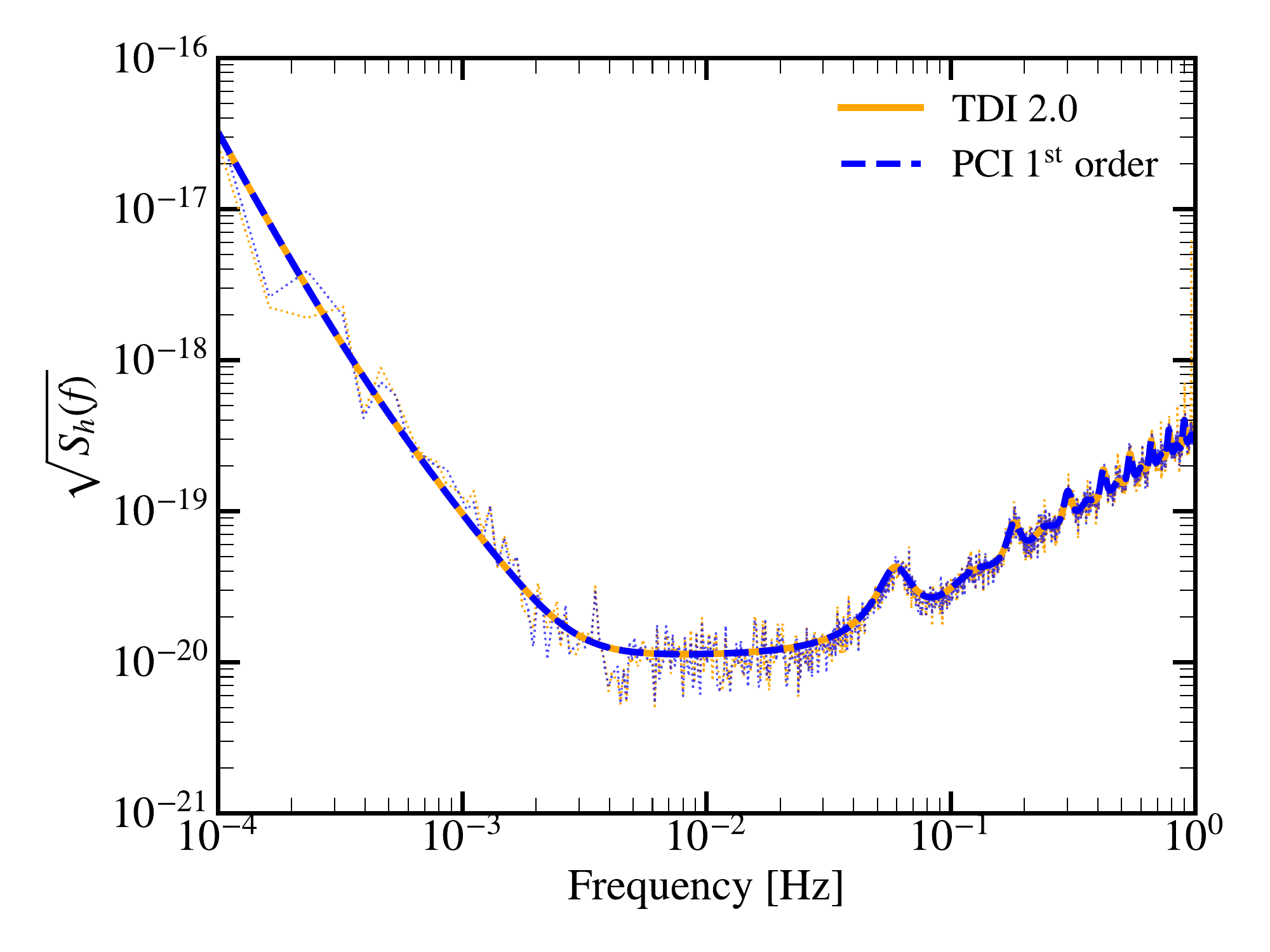}
    \caption{Sky-averaged sensitivities of first-order aPCI (dashed blue) and second-generation TDI (orange). The thick solid curves are the analytical sensitivities and the dotted thin lines are empirical estimates of the sensitivity obtained with GW stochastic background and instrumental noise simulations, and computing the ratios of their smoothed periodogram.}
    \label{fig:sensitivities}
\end{figure}

We plot the first-order aPCI total sensitivity with a thick, dashed blue line in Fig.~\ref{fig:sensitivities} thanks to the analytical model provided by Eq.~\eqref{eq:total_sensitivity}. As a comparison, we do the same for the case of  second-generation TDI with a solid orange line. The aPCI and TDI curves match remarkably well, given that aPCI does not use any model describing the laser frequency noise terms appearing in the link measurements, nor any prior knowledge of light travel time delays between the spacecrafts. The computation of noise-cancelling combinations directly comes from the singular spectrum analysis of one realization of the laser-frequency-noise dominated data matrix $\bm{Z}^{(1)}$, from which we obtained the singular vectors $\bm{V}^{(1)}$.

To check that this analytical results represent what is actually measured, we plot the empirical sensitivities defined as the ratios of the response periodograms of Fig.~\ref{fig:aPCI_gw_response} and the noise periodograms of Fig.~\ref{fig:aPCI_noise_models},
\begin{equation}
\label{eq:total_estimated_sensitivity}
    \hat{S}_{h, e_{\perp}}(f) = \left[ \sum_{j=1}^{3} \frac{\lvert \tilde{e}_{\perp, \mathrm{GW} j} \rvert^{2}}{\lvert \tilde{e}_{\perp, \mathrm{n} j} \rvert^{2}} \right]^{-1}.
\end{equation}
The above equation only uses the outputs of our GW background and instrumental noise simulations. These quantities are empirical estimates of each aPCI variable's SNRs.
The plot shows that the analytical models are consistent with their empirical equivalents. The aPCI processing is therefore a valid method for practical data analysis purposes.
 
\begin{figure}[!h]
    \centering
    \includegraphics[width=\columnwidth, trim={0.5cm, 0.3cm, 0.5cm, 0.3cm}, clip]{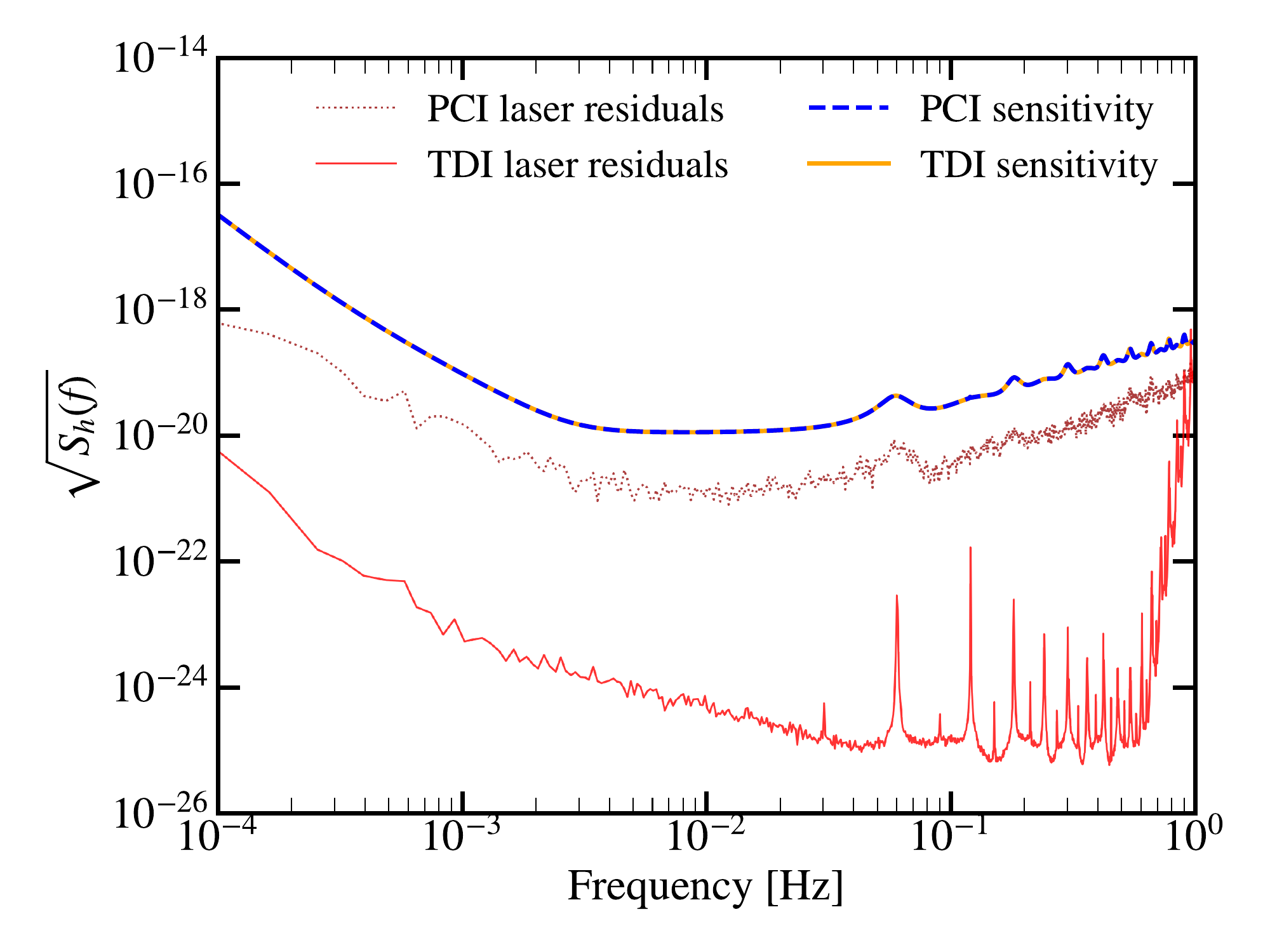}
    \caption{Residual laser noise in the aPCI (dotted brown) and TDI (solid red) sensitivities, compared with their total sensitivity (dashed blue and orange curves).}
    \label{fig:sensitivities_with_laser_noise}
\end{figure}

Unlike in Paper I, we approximated the aPCI residual noise covariance matrix $\bm{\tilde{\Sigma}}_{{e}_{n}}$ by only accounting for secondary noises, and zeroth order effects in time. From this approximation, we derived the matrix's eigenvectors $\bm{\Phi}$ that yield the orthogonal variables $\bm{\tilde{e}}_{\perp}^{(1)}$. In reality, a non-zero residual laser noise is still present in the aPCI variables, which slightly modifies their covariance, and hereby their eigen-decomposition. To see that, we set up a simulation including only laser frequency noise, discarding all other noise sources. Then we apply the exact same aPCI decomposition to evaluate the level of laser noise that remains. We show the outcome in Fig.~\ref{fig:sensitivities_with_laser_noise}.

For comparison, in the figure we reproduce the total sensitivity of both aPCI (dashed blue curve) and TDI (solid orange curve). We draw in dotted brown lines the \hl{strain noise in the aPCI variables due to laser frequency residuals. We compute it with Eq.~}\refeq{eq:total_estimated_sensitivity} \hl{where the variables $\tilde{e}_{\perp, \mathrm{n} j}$ are now the outputs of the aPCI decomposition from the laser-noise-only simulation.} The curve stands about one order of magnitude below the total sensitivity. Thus, as we already observed in Fig.~\ref{fig:aPCI_noise_periodograms}, the aPCI algorithm suppresses laser frequency noise \hl{two orders of magnitude} below the level of secondary noises in power. The same residuals in the TDI variables (plotted in solid red line) are down to about \hl{six} orders of magnitude \hl{in power} below all other noises in most of the frequency band. \hl{This difference is due to the fact that TDI is explicitly designed to cancel laser noise, whereas aPCI is rather constructed to find the data combinations minimizing the variance. Furthermore, here the TDI residuals are obtained in an idealized case where the inter-spacecraft delays are perfectly known, which will not be true in practice. A fairer comparison could be done with TDI-ranging (TDIR)}~\cite{tinto_time-delay_2005, page_bayesian_2021} \hl{where light travel time delays are estimated from the data using standard TDI combinations, yielding residuals at least $5 \times 10^3$ lower than secondary noise.}

That said, part of the residual difference we observe may be due to the non-optimal orthogonalization. More accurately characterizing the noise in the aPCI variables $\bm{\tilde{e}}$ may help further decreasing the aPCI laser frequency noise residuals. We leave this task to future work, as Fig.~\ref{fig:sensitivities_with_laser_noise} already demonstrates that, in its current form, the aPCI approach matches the TDI sensitivity up to a relatively small error. Using the dotted brown residual noise periodogram, we estimate this error to be 2\% on average.

\section{\label{sec:discussion}Discussion}

We further developed the data-driven technique for space-based interferometry introduced in our earlier work (Paper I). In the present study, we allowed the distance in-between the spacecraft to vary with time, breaking the stationarity of the laser noise that dominates interferometer measurements. To make our aPCI method suitable for such a time-varying configuration, we included a new time-dependent term in the data matrix that we analyze through singular value decomposition. Instead of carefully modelling \hl{the laser phase contributions}, the aPCI approach ``learns" the filter coefficients that one needs to apply to the data to mitigate laser frequency noise.

We showed that this extension allows us to cancel laser noise down to a level 100 times better than the previous version which worked in the case of constant arm lengths. This level is enough to reduce laser frequency noise residuals below the other noise sources. Based on an approximate model for the residual noise covariance, we transformed the aPCI components into a quasi-orthogonal set of variables of which only three are GW-sensitive. We demonstrated that their combined sensitivity is the same as for second-generation TDI, up to a 2 \% relative error. This result shows that we can infer all the information needed to process space-based interferometry data without a particular model \hl{describing how laser noise enters the measurements or what is its covariance}. The only implicit assumption is that the data features significant time correlations over a specific duration. Extracting these correlations via \hl{singular spectrum analysis} leads to a data-driven basis where we can separate the laser-noise-free components of the data from the laser-noise-dominated ones.

We find that the aPCI laser frequency noise residuals lie \hl{two orders of magnitude} below the level of secondary noises (in power spectral density), which allows one to use the methods for most GW data analysis purposes. Nevertheless, in the present implementation, these residuals stand higher than the level achieved by second-generation TDI \hl{with exact inter-spacecraft delays}. The reason is that \hl{aPCI infers the underlying noise correlations from the data up to a certain statistical precision}. In addition, after getting the lowest-variance aPCI variables, we need to orthogonalize them with respect to the remaining noise, in the same way we construct the optimal TDI combinations A, E, and T from the Michelson combinations. Similarly as in TDI, this orthogonalization process requires knowing the variables' covariance matrix. In this work, we computed this covariance using an approximation which only includes non-laser-frequency-noise contributions. This way, the covariance is easy to derive from both the single-link noise PSDs and the singular vectors and does not require any knowledge of the laser noise correlations structure. However, ignoring them leads to a slightly imperfect diagonalization of the covariance, which, in turn, yields a set of three variables with non-optimal sensitivity. To reach the full potential of aPCI (but also of TDI), we would therefore need to characterize the aPCI variables' residual noise covariance from the data. \hl{This would allow us to account not only for any laser noise residuals, but also for secondary noise's features such as unequal PSD levels and correlations among link measurements.} We plan to develop this characterization in further works. 


In conclusion, the present form of the aPCI method is already operational for practical purposes. \hl{It provides a complementary approach to classic TDI that could help validating the noise reduction pipeline, which is a critical step for space-based GW observation.} Besides, further improving aPCI's sensitivity is possible, provided that we develop a robust frequency-domain covariance estimator. We also plan to focus on understanding how aPCI's sensitivity depends on its tuning parameters. Indeed, while not critical, a trade-off between performance and computational efficiency most likely exists when choosing the analyzed data size $N$, the stencil size $n_h$, the number of components $q$ to consider, and the order $p$ of the Taylor expansion in time. We will therefore assess the influence of these parameters in further studies. Furthermore, testing more realistic configurations including additional noises and laser locking is required~\cite{tinto_implementation_2003, sylvestre_locking_2004}. \hl{Injecting various GW sources would also allow us to test the algorithm's robustness against the presence of signals, although a preliminary assessment in Paper I suggests that it would take unrealistically high SNRs to cause any significant impact}. Finally, we envision to demonstrate Bayesian inference of GW source parameters with the aPCI framework, which would be the ultimate demonstration of its reliability for GW data analysis.

\appendix

\section{\label{sec:secondary_noises}Secondary noises model}
In this section we write down the analytical expressions for the acceleration and OMS noises PSDs forming the secondary noises in Eq.~\eqref{eq:secondary_noises_psd}.
The acceleration noise PSD is
\begin{eqnarray}
S_{\mathrm{a}}(f) = \left(\frac{a_{\mathrm{TM}}}{2\pi c f} \right)^2 \left(1 + \left(\frac{f_{-2}}{f}\right)^2\right) \left(1 + \left(\frac{f}{f_4}\right)^4\right),
\end{eqnarray}
where the PSD is expressed in fractional frequency deviations (FFD) and the level of acceleration noise is $a_{\mathrm{TM}} = 2.4 \times 10^{-15} \, \mathrm{ms^{-2}Hz^{-1/2}}$. The pivot frequencies are set to $f_{-2} = 0.4$ mHz and $f_{4} = 8$ mHz. 
Besides, we allocate the OMS noise to 
\begin{eqnarray}
S_{\mathrm{OMS}}(f) = a_{\mathrm{OMS}}^{2}  \left( \frac{2\pi f}{c} \right)^2 \left(1 + \left( \frac{f_{-4}}{f} \right)^4 \right),
\end{eqnarray}
with a level of $a_{\mathrm{OMS}} = 6.14 \times 10^{-12} \, \mathrm{m Hz^{-1/2}}$ which corresponds to the noise affecting the science interferometer, \hl{and a pivot frequency of $f_{-4} = 2 \times 10^{-3}$ Hz}. Noises coming from other interferometers are ignored in this study.

\section{\label{eq:average_operator}Definition of the power spectrum operator}
Throughout the paper, we define the continuous-time cross power spectrum of two continuous time functions $u(t)$ and $v(t)$ as
\begin{equation}
\langle \tilde{u}(f) \tilde{v}^{\ast}(f') \rangle = \lim_{T \to +\infty} \frac{1}{T} \operatorname{E}[\tilde{u}_{T}(f) \tilde{v}^{\ast}_{T}(f')],
\end{equation}
where $\tilde{u}_{T}$ is the continuous-time Fourier transform of $\tilde{u}$ when observed on a finite duration $T$:
\begin{equation}
\tilde{u}_{T}(f) \equiv \int_{-\frac{T}{2}}^{+\frac{T}{2}} u(t) e^{-2 \pi i f t} dt.
\end{equation}

\begin{acknowledgments}
 We would like to thank Nicolas Le Bihan from GIPSA-Lab, for his very useful feedback and insights on singular spectrum analysis. Many thanks to the NASA LISA Study Office team for fruitful discussions when developing the PCI framework. We also thank Jean-Baptiste Bayle and Martin Staab and their collaborators for making their codes available. We would also like to thank the referees for their thorough review which helped improving the manuscript. This work was performed using HPC resources from CNES Computing Center.
\end{acknowledgments}

\bibliography{library}

\end{document}